\documentclass[prb]{revtex4}
\usepackage{epsf}
\begin{document}
\title{Three Faces of the Aharonov-Bohm Phase}
\author{Patrick Das Gupta}
\affiliation{Department of Physics and Astrophysics, University of Delhi, Delhi - 110 007 (India)}
\email{pdasgupta@physics.du.ac.in}

\begin{abstract}

Beginning with the basic notions of quantum theory,  impossibility of  `trajectory' description for particles that  ensues from uncertainty principle is discussed.  Why the observed tracks in bubble/cloud chambers are not really the `trajectories' of high energy particles, rather they are simply the trails of the atoms/molecules excited or ionized in the direction of the high momenta of the incoming particles, are  highlighted. 

Thereafter, the notion of symmetry and its application to the non-relativistic Schrodinger equation have  been delineated. The demand for U(1) gauge invariance and the resulting `gauge miracle', that automatically  leads to the correct interaction terms describing the electromagnetic force between the charge particles and the field, have been elaborated upon.

A  simple treatment, but with explicit derivation,  of the Aharonov-Bohm effect, is presented, that   underlines a strange phenomena - sites, impenetrable by charge particles but  are threaded  with magnetic field lines,  conspire with a `Trojan Horse', namely the magnetic vector potential outside,  to cause  measurable shift in the double-slit interference fringes. This shift in the interference pattern is non-classical since the charge particles move in field-free regions. 

The crucial Aharonov-Bohm (AB) phase that makes its entry in the above bizarre effect is also deployed to derive the observed magnetic flux quantisation in superconductors as well as the Dirac result which implies that the  existence of a single magnetic monopole anywhere in the universe would entail quantisation of the product of  a particle's electric charge and  the monopole's magnetic charge. Nontrivial consequences of AB phase follow whenever the physical region accessible, from the point of view of the wavefunction of a charge particle, is multiply-connected.

\end{abstract}


\maketitle

\section{INTRODUCTION}
Our world is fundamentally quantum mechanical in nature. We would not have existed otherwise. This is because of the following reason: Matter is made of  atoms that consist of negatively charged electrons hovering around the positively charged nuclei, bound by their mutual electrostatic attraction. Now, classical mechanics says that any particle that goes around a point is necessarily  accelerating. (In the atom's case, the acceleration is due to the Coulomb force between the electrons and the nucleus.)

 Also, classical electromagnetic theory  tells us that an accelerating charge always radiates  electromagnetic waves, that carry away energy from the system of charge particles in motion. But this would imply  that atomic  electrons, as they lose their orbital energy by emitting  electromagnetic radiation, would spiral into the nuclei in about $10^{-8}$ s, making all atoms unstable! But then, how is it that we and the world   exist?

Quantum mechanics (QM) resolves this paradox by adopting a radical stance: In atoms and molecules, the electrons do not orbit around their respective nuclei, nor do they accelerate and thus, there is no question of their  emitting electromagnetic waves incessantly. Instead, electrons in atoms/molecules occupy discrete and very specific energy levels that follow from quantum rules. Hence, atoms and molecules in their respective ground states are stable. And, so is the universe. 

But if the electrons are not flying around the nuclei, what kind of trajectories do they follow in  atoms? Now, in order to completely determine  the  future trajectory $\vec{r}(t) $ of any particle in Newtonian  dynamics, one needs    to solve,
\begin{equation}
m \frac {d^2 \vec{r}} {dt^2}   \equiv\ m \ddot{\vec{r}}= \vec{F}(\vec{r},t)\ ,
\end{equation}
 where $\vec{F}(\vec{r},t)$ is  the force acting on the particle of mass $m$ when it is at position $\vec{r}$, at time $t$. However, to obtain a particular solution of eq.(1), it is necessary to   specify a precise initial  position vector $\vec{r}(t_i)$ as well as a precise velocity $\vec{v}(t_i)$  $\equiv\ \dot{\vec{r}} \vert_{t_i}$ at some initial time $t_i$. 
 
On the other hand, in QM,  Heisenberg's {\bf uncertainty principle} (HUP)  states that, no matter what, it is impossible to determine with arbitrary precision the values of position and momentum (or, equivalently, velocity $\equiv$ momentum/mass) simultaneously, for any particle. Therefore, HUP  precludes one to provide both $\vec{r} (t_i)$ and $\vec{v} (t_i)$ at time $t_i$, with very high accuracy, prohibiting one to derive a solution of eq.(1).  

Hence,  trajectories as physical description of particle states is meaningless in quantum theory. For  otherwise, given a trajectory $\vec{r}(t)$ one can always determine the velocity at any instant $t$ by determining $ {d \vec{r}}/ {dt}$, violating thereby  the uncertainty principle. HUP is essential to QM, and has been vindicated time and again in all experiments conducted so far. It is the classical observables like velocity, acceleration and force that are not fundamentally meaningful in QM.  

Instead, QM postulates that for every physical state of any non-relativistic particle there exists a complex-valued   wavefunction $\psi(\vec{r},t)$ which provides a complete description of the particle's state. Such a wavefunction  can  always be expressed as,
\begin{equation}
\psi(\vec{r},t)= R(\vec{r},t)\exp{(i S (\vec{r},t))}
\end{equation}
 because of the Euler formula $\exp{(i\theta)}= \cos{\theta} + i \sin{\theta}$, where $R(\vec{r},t)$ and $ S (\vec{r},t)$ are two real-valued functions.
 
For a non-relativistic particle, the physical meaning of $\psi(\vec{r},t)$  is gleaned from the Born rule, according to which, $\vert \psi(\vec{r},t) \vert ^2 \ d^3r$ is the probability of finding the particle at time $t$ in a tiny volume element $d^3r$ located at  $\vec{r}$. The phase information represented by the real function $S (\vec{r},t)$ in eq.(2) is irrelevant  when one considers the probability  density $\vert \psi(\vec{r},t) \vert ^2 = R^2(\vec{r},t)$. But in interference experiments, as we shall see later, $S (\vec{r},t)$ plays a crucial role. (For an alternative  interpretation of  meaning of the wavefunction,  the readers are referred to the articles [1]-[3].)
 
 In atoms, the electrons are described by stationary  wavefunctions that are  eigenfunctions of the Hamiltonian operator $\hat{H}$ (i.e. a linear and hermitian  operator corresponding to the energy observable). Only the phase $S(\vec{r},t)$ of a stationary wavefunction changes with time, and hence, electrons  occupying the energy levels of atoms that are not perturbed in any manner, remain `stay put' in those levels, as $R(\vec{r})$ is independent of time. Therefore, quantum systems occupying their lowest energy levels (i.e. ground states) are  stable. (Electrons constantly face randomly varying electric and magnetic fields arising out of quantum vacuum fluctuations and thus, those in the higher energy states  make spontaneous transitions to lower energy levels.)

One may wonder  if indeed HUP  prevents  particles to have well defined trajectories then what on earth are the paths that one sees in bubble chambers (or cloud chambers, scintillating detectors, etc.)  when  high energy particles (produced in accelerators or those associated with cosmic rays) pass through them? Actually, an observed track in such a situation is  simply  an array of cluster of atoms/molecules in the detector that have got excited or ionized due to the energy exchanges with the trespassing elementary particle.

 Since a high energy particle possesses a large linear momentum, after it interacts and excites a particular atom/molecule  (which in turn, imparts momentum and energy to other atoms/molecules surrounding it, leading to their  excitations) present in the detector, the probability of interacting with the next atom/molecule is higher when the latter lies in the direction of the incident particle's momentum \cite{bec3}. This process goes on as long as the intruder particle is within the bubble or cloud chamber,  and continues to have very high energy.
 
 This `forward' favouring interactions lead to linear  paths traced out by  groups of excited atoms/molecules lying in the direction of the incoming high energy particles.  Therefore, a track in a bubble chamber or a detector is just  a macroscopic collection of excited/ionized atoms  corresponding to the sequential excitation  along  the direction of the momentum of the incident high energy particle, and is certainly not the intruder particle's non-existent intrinsic path. 

\section{Wavefunction, Complex numbers, Measurement and Chance}

The time evolution of $\psi(\vec{r},t) $, the wavefunction encountered in the previous section,  is determined by the {\bf Schrodinger equation} (please refer to the chapter by Professor Ajoy Ghatak in this  volume \cite {bec4}),
\begin{equation}
i \hbar {{\partial \psi}\over{\partial t}} = - \frac {\hbar^2}{2m} \nabla ^2 \psi   + V (\vec {r}, t)\psi  
\end{equation}
where $V (\vec {r}, t)$ is the potential energy of the particle. What eq.(1) is to classical mechanics, eq.(3) is to  QM, i.e. the dynamics in QM is governed by eq.(3).
 
There are three  remarkable features of the Schrodinger equation: (a) If one knows the initial state $\psi(\vec{r},t_0) $ at the instant $t_0$, one can determine the future state $\psi(\vec{r},t) $ for $t> t_0$  by solving eq.(3), (b) if $\psi_1(\vec{r},t) $ and $\psi_2(\vec{r},t) $ are two independent solutions of eq.(3) then $C_1 \psi_1(\vec{r},t) + C_2 \psi_2(\vec{r},t) $ is also a solution, where $C_1$ and $C_2$ are arbitrary complex numbers (since eq.(3) is a linear differential equation), and (c) the imaginary number $i$ and the reduced Planck's constant $\hbar $ appear explicitly in this dynamical equation.

Property (a)  tells us that, according to the Schrodinger equation, the physical state changes with time in a causal fashion with randomness or probabilities playing no role whatsoever in its time evolution. The linearity feature (b) leads to  wave-like  behavior  of particles demonstrated  in two-hole/slit  experiments whereby interference patterns are observed  because of linear superposition of wavefunctions (please refer to the chapter by Professor Ajoy Ghatak \cite {bec4}).

 Also, because of (b), $\psi (\vec{r},t) $ and $\psi^\prime (\vec{r},t) = e^{i  \alpha} \psi (\vec{r},t) $ cannot be distinguished physically from each other, as long as $\alpha $ is an arbitrary and  real constant.  According to (c), complex numbers are as fundamental to the physical world as real numbers. Hence,  $i=\sqrt{-1}$ is not just a mathematical device that is invented to solve the quadratic equation $x^2+1=0$, but is very much part of the real world. An interesting  aspect in this regard is that $i$ and $\hbar $ seem to invariably appear  together in the quantum theory.

In QM, the wavefunction, $\psi(\vec{r},t) $,  provides a complete description of the state of a non-relativistic particle that has $0$ spin. For a  particle having an intrinsic spin, the state is specified  by a column vector of wavefunctions.  It was Max Born who realized first that $ \vert \psi (\vec{r},t)\vert ^2$ gives the probability density associated with the measurement of a particle's position at time $t$. \cite{bec5} This  was briefly alluded to, in section I, while describing the physical significance of the wavefunction. 
 
So, one cannot   guess with certainty the outcome of a position measurement on a spin $0$, non-relativistic particle that is described by a wavefunction $\psi(\vec{r},t) $. However, one can predict the  chance or probability $P(\vec {r},t)$ of finding the particle  within an infinitesimal volume-element $dV$ surrounding the position vector  $\vec {r}$ at time $t$ to be,
\begin{equation}
P(\vec {r},t)= \vert \psi (\vec{r},t)\vert ^2 dV
\end{equation}

One may ask, what is the source of this randomness since the time evolution of the state of a particle is causal, as it is completely determined by solving eq.(3) given a potential energy $V(\vec{r},t)$ that describes the interaction of the particle with an external system. After all,  a  measurement of the position  (or, for that matter any other observable) is nothing but an interaction between the particle and the measuring apparatus (like a bubble chamber, for example), and this in principle should be described by eq.(3). 

The problem of indeterminism in QM, intimately tied with the source of the Born rule given by eq.(4), has not 
fully been resolved in a satisfactory manner. Although there are several excellent models that attempt to explain the appearance of an {\it ad hoc} rule like eq.(4) while maintaining a deterministic dynamical evolution of the quantum state (see, e.g. [7], [8] and [9]).

\section{Symmetry, Phase of a Wavefunction, Electromagnetism and All That}

The renowned mathematician Hermann Weyl had defined a {\bf symmetry} associated with a given  entity to be an operation on the latter that leaves the said entity  indistinguishable from its previous state. For instance, if someone rotates a smooth and ideal  sphere about an axis passing through its centre by any amount when we are not present, then we cannot figure out whether such an action really did take place in our absence. Hence, a collection of arbitrary rotations form a symmetry as far as the sphere is concerned.

Now, in the previous section, it was pointed out that the Schrodinger equation, given by eq.(3), is form invariant under the constant phase transformation $\psi (\vec{r},t) \ \rightarrow$ $\psi^\prime (\vec{r},t) = e^{i  \alpha} \psi (\vec{r},t) $. Therefore, such a transformation is a global symmetry (`global' because the phase $\alpha $ does not depend on space-time coordinates) of the Schrodinger equation.

Similarly, in classical mechanics, the trajectories of a particle do not get modified if one adds an arbitary constant $V_0$ to its  potential energy,  since the force is given by the gradient of the potential energy. Hence, a constant shift of the zero of the potential energy has no consequences on the particle motion, and thus is a symmetry. What happens if such a shift is performed in QM?

Consider the transformation  $\psi (\vec{r},t) \ \rightarrow$ $\psi^\prime (\vec{r},t) = e^{-\ i V_0 t/\hbar} \psi (\vec{r},t) $ so that,
\begin{equation}
i \hbar \frac {\partial \psi^\prime (\vec{r},t)}{\partial t} = i \hbar \bigg ( - \frac {i V_0} {\hbar} e^{-\ i V_0 t/\hbar} \psi (\vec{r},t) + e^{-\ i V_0 t/\hbar} \ \frac {\partial \psi (\vec{r},t)} {\partial t} \bigg ) = V_0 \psi^\prime (\vec{r},t) + e^{-\ i V_0 t/\hbar} \bigg ( - \frac {\hbar^2}{2m} \nabla ^2 \psi   + V (\vec {r}, t)\psi \bigg )
\end{equation} 
$$=  - \frac {\hbar^2}{2m} \nabla ^2 \psi^\prime (\vec{r},t)    + [V (\vec {r}, t) + V_0] \psi^\prime (\vec{r},t)  \ .$$ 
Hence, the above calculation shows that the combination of adding an arbitrary real constant $V_0$ to the old potential energy along with multiplying a phase factor $e^{-\ i V_0 t/\hbar}$  to the old wavefunction is a symmetry of the Schrodinger equation. The  probabilities that can be inferred from experiments remain unchanged since $\vert \psi^\prime (\vec{r},t)\vert ^2 = \vert \psi (\vec{r},t)\vert ^2 $, in this case of a transformation in which the phase depends linearly on time.

Let us  delve into another kind of transformation  in which the phase $\alpha $ now depends on both space and time. Such transformations are  referred to as the {\bf gauge} transformations. So, consider the following operation,
\begin{equation}
\psi (\vec{r},t) \ \rightarrow \ \psi^\prime (\vec{r},t) = e^{i \frac {q \chi(\vec{r},t)} {\hbar c} } \psi (\vec{r},t)
\end{equation}
where $q$ is a real constant (which may be identified as the electric charge, as we shall see later),  $c$ is the speed of light in vacuum and $\chi(\vec{r},t) $ is an arbitrary but a differentiable, real function of space and time. The derivatives of the new wavefunction are related to that of the old wavefunction in the following way,
\begin{equation}
\nabla \psi^\prime = e^{i \frac {q  \chi(\vec{r},t)} {\hbar c} } (\nabla \psi + i \frac {q} {\hbar c}\  \psi \nabla \chi) \ \ \mbox{and}\ \ \frac {\partial \psi^\prime}{\partial t} = e^{i \frac {q  \chi(\vec{r},t)} {\hbar c}} \bigg (\frac {\partial \psi}{\partial t} + i\frac {q} {\hbar c}\  \psi \frac {\partial \chi}{\partial t} \bigg )
\end{equation}
Because of the presence of derivatives of $\chi $ in  eq.(7), it is obvious  that the Schrodinger equation is not form invariant under the gauge transformation given by eq.(6). But then, it also reminds us of a transformation of potentials that we frequently encounter in electromagnetic (EM) theory.
 
In classical electromagnetism, it is often useful to express electric and magnetic fields  in terms of the potentials $\phi (\vec{r},t)$ and $\vec{A} (\vec{r},t)$, 
\begin{equation}
\vec{E} (\vec{r},t)= - \nabla \phi - \frac {1} {c} \frac {\vec{A} (\vec{r},t)} {\partial t} \ \ \ \mbox{and} \ \ \ \vec{B} (\vec{r},t) = \nabla \times \vec{A} \ \ \ .
\end{equation}
 
One can always measure the EM fields $\vec {E}$ and $\vec {B}$ from their actions on a test charge having mass $m$, charge $q$ and a non-relativistic velocity $\vec {v}$, using the Lorentz force equation,
\begin{equation}
m \frac {d \vec{v}} {dt} = q \bigg (\vec{E} (\vec{r},t) + \frac {\vec {v}} {c} \times \vec {B}(\vec{r},t) \bigg ) =  q \bigg (\vec{E} (\vec{r},t) + \frac {\dot {\vec {r}}} {c} \times \vec {B}(\vec{r},t) \bigg ) \ \ ,
\end{equation}
where $\dot {\vec {r}} \equiv \vec {v}$

On the other hand, the potentials $\phi (\vec{r},t)$ and $\vec{A} (\vec{r},t)$ are not themselves  observable, for they are not unique. Given an arbitrary smooth and real function $\chi (\vec{r},t)$,  it can be easily shown using eq.(8) that under the following transformations,

\begin{equation}
\vec{A}(\vec{r},t) \rightarrow \vec{A}^\prime (\vec{r},t)=\vec{A}(\vec{r},t) + \nabla \chi
\end{equation}
and,
\begin{equation}
\phi (\vec{r},t) \rightarrow \phi^\prime (\vec{r},t) = \phi (\vec{r},t) - \frac {1}{c} \frac {\partial \chi}{\partial t} \ ,
\end{equation}
 the $\vec {E}$ and $\vec {B}$ fields are invariant. Eqs.(10) and (11) are referred to as  the gauge transformation of the EM fields.
 
 Comparing eqs.(7), (10) and (11) we can guess  the following dynamical equation for the wavefunction of a charge particle,
 \begin{equation}
 i \hbar {{\partial \psi}\over{\partial t}} = \frac {1} {2 m} \bigg (- i \hbar \nabla - \frac {q} {c}\  \vec{A}(\vec{r},t)\bigg ). \bigg (- i \hbar \nabla - \frac {q} {c}\  \vec{A}(\vec{r},t) \bigg ) \psi   + q\ \phi (\vec {r}, t)\ \psi  
\end{equation}
$$= - \frac {\hbar^2} {2 m} \bigg (\nabla - \frac {i q} {\hbar c}\  \vec{A}(\vec{r},t)\bigg ). \bigg ( \nabla - \frac {i q} {\hbar c}\  \vec{A}(\vec{r},t) \bigg ) \psi   + q\ \phi (\vec {r}, t)\ \psi  \ \ ,$$
that exhibits a gauge symmetry.

One can easily prove by making use of eqs.(7), (10) and (11) that,
\begin{equation}
\nabla \psi^\prime - i \frac {q} {\hbar c} \vec {A}^\prime \psi^\prime = e^{i \frac {q  \chi(\vec{r},t)} {\hbar c} } \bigg (\nabla \psi - i \frac {q} {\hbar c} \vec {A} \psi \bigg )
\end{equation}
\begin{equation}
\nabla . \bigg (\nabla \psi^\prime - i \frac {q} {\hbar c} \vec {A}^\prime \psi^\prime \bigg ) = e^{i \frac {q   \chi(\vec{r},t)} {\hbar c} } \bigg ( \nabla +  i \frac {q} {\hbar c} \nabla \chi \bigg ) . \bigg (\nabla \psi - i \frac {q} {\hbar c} \vec {A} \psi \bigg )
\end{equation}
Employing eqs.(7), (13) and (14) in  eq.(12), one  immediately derives the result,
\begin{equation}
 i \hbar {{\partial \psi^\prime}\over{\partial t}} = \frac {1} {2 m} \bigg (- i \hbar \nabla - \frac {q} {c}\  \vec{A}^\prime(\vec{r},t)\bigg ). \bigg (- i \hbar \nabla - \frac {q} {c}\  \vec{A}^\prime(\vec{r},t) \bigg ) \psi^\prime   + q\ \phi^\prime (\vec {r}, t)\ \psi^\prime  
\end{equation}
which demonstrates that the  Schrodinger equation (eq.(12)) is form invariant under the full gauge transformation  given by the combination of eqs.(6), (10) and (11). Hence, the gauge transformation is a symmetry of this dynamical theory - all the physical consequences remain  the same whether one uses the new wavefunction and the new EM potentials or the old ones.  

The above analysis points to  a `gauge miracle', in the sense that a demand for gauge invariance (i.e. symmetry associated with the local or space-time dependent phase $\chi (\vec {r}, t)$) of the Schrodinger equation automatically leads to the correct quantum theory of charge particles interacting with EM fields. Incidentally, the quantum field theories of strong and electro-weak interactions (that have been verified   in accelerator experiments numerous times) were arrived at, using essentially the principle of  gauge invariance.

\section{The Aharonov-Bohm Phase, Superconductors and Magnetic monopoles}

In the discussion so far, we have witnessed that the wavefunction $\psi $ as well as the EM potentials $\phi $ and  $\vec{A}$ are themselves not measurable directly, although physical quantities derived from 
them like probabilities and EM fields are. So, are $\psi $, $\phi $ and  $\vec{A}$  just mathematical book-keeping tools? In section I, I had remarked that it was QM which elevated the role of $\sqrt{-1}$ from being just a mathematically beautiful but  shadowy  idea to being an actual indispensable actor in  the real world.

 Yakir Aharonov and David Bohm, in 1959, did the same for the  EM potentials and showed that regions forbidden to matter can conspire with the EM potentials to display surprising effects that are directly measurable \cite{bec9}. Following their work,  we now consider the dynamics of a non-relativistic particle having an electric charge $q$  in the presence of only a static magnetic field $\vec{B} (\vec{r})$. Then, from eq.(12), the wavefunction describing the state of such a particle will satisfy,
\begin{equation}
i \hbar {{\partial \psi}\over{\partial t}} = \frac {1} {2 m} \bigg (- i \hbar \nabla - \frac {q} {c}\  \vec{A}(\vec{r})\bigg ). \bigg (- i \hbar \nabla - \frac {q} {c}\  \vec{A}(\vec{r}) \bigg ) \psi  
\end{equation}
because of the fact that $\vec {E}=0$ as well as  time-independent $\vec {B}$ necessitates $\frac {\partial \vec{A}}  {\partial t}$ to vanish and $\phi $ to be an arbitrary constant (which has no physical significance as discussed in the beginning of section III) that can be set to $0$ without any loss of generality. 

Suppose, we already know the solution $\psi_0 (\vec{r},t) $ of the free particle Schrodinger equation,
\begin{equation}
i \hbar {{\partial \psi_0}\over{\partial t}} =  - \frac {\hbar^2}{2m} \nabla ^2 \psi_0 \ \ .
\end{equation}
Then, can we say something about the solution $\psi (\vec{r},t) $  of eq.(16)? The answer is a big YES if we are interested in the case when the charge particle is confined to regions where $\vec {B}=0$, even though $\vec {B} \neq 0$ somewhere else that is not accessible to the test charge.

The  solution, as we shall see shortly, in such situations  is given by,
\begin{equation}
\psi(\vec{r},t)=e^{\frac {i q}{\hbar c}\int^{\vec{r}}_{\vec{r}_0 \ C} {\vec{A}(\vec{r'}).d\vec{r'}}}\psi_0 (\vec{r},t)
\end{equation}
where $d\vec{r'}$ is an infinitesimal  line-element tangent to a curve  $C$ at $\vec {r'}$ that lies only in the region where 
$\vec {B}=0$, beginning from $\vec {r}_0$ and terminating at the position vector $\vec {r}$.

An alert reader will protest stating that the RHS of eq.(18) is path dependent and so, how can  it be a valid solution? That is {\bf incorrect} because in the regions   where $\vec {B}=\nabla \times \vec {A}=0$, we may express the vector potential as $\vec {A} (\vec{r}) = \nabla \zeta (\vec{r})$ where $\zeta (\vec{r})$ is a real, differentiable function. In that case,
\begin{equation}
\int^{\vec{r}}_{\vec{r}_0} {\vec{A}(\vec{r'}).d\vec{r'}}= \int^{\vec{r}}_{\vec{r}_0} {\nabla \zeta(\vec{r'}).d\vec{r'}} = \zeta (\vec{r}) - \zeta (\vec{r}_0)
\end{equation}
is {\bf path independent} as long as the path lies in the field free region. Hence, the wavefunction given by eq.(18) is well defined and single-valued. 

Let us now prove that the wavefunction given by eq.(18) indeed satisfies eq.(16) in places where $\vec{B}=0$. Since the phase factor appearing in eq.(18) is path independent in the field free regions, it is a single-valued function of $\vec{r}$ and therefore,
\begin{equation}
\nabla \psi = e^{i \frac {q} {\hbar c}\int^{\vec{r}} {\vec{A}(\vec{r'}).d\vec{r'}} } \nabla \psi_0 + i \frac {q} {\hbar c} \vec{A} (\vec{r}) e^{i \frac {q} {\hbar c} \int^{\vec{r}} {\vec{A}(\vec{r'}).d\vec{r'}}} \psi_0 \ \ ,
\end{equation}
so that,
\begin{equation}
\bigg (\nabla - i \frac {q} {\hbar c} \vec{A} (\vec{r}) \bigg) \psi = e^{i \frac {q} {\hbar c}\int^{\vec{r}} {\vec{A}(\vec{r'}).d\vec{r'}} } \nabla \psi_0 \ \ .
\end{equation}
Applying the operator $\nabla - i \frac {q} {\hbar c} \vec{A}$ once more, by taking the dot product, to the above equation leads to,
\begin{equation}
\bigg (\nabla - i \frac {q} {\hbar c} \vec{A} \bigg ). \bigg (\nabla - i \frac {q} {\hbar c} \vec{A} \bigg ) \psi = \nabla . \bigg ( e^{i \frac {q} {\hbar c}\int^{\vec{r}} {\vec{A}(\vec{r'}).d\vec{r'}} } \nabla \psi_0 \bigg ) - i \frac {q} {\hbar c} \vec{A} . \bigg (\nabla - i \frac {q} {\hbar c} \vec{A} (\vec{r}) \bigg) \psi
\end{equation} 

\begin{equation}
 =  e^{i \frac {q} {\hbar c}\int^{\vec{r}} {\vec{A}(\vec{r'}).d\vec{r'}} }\bigg [ \bigg (i \frac {q} {\hbar c}  \bigg) \vec{A}.\nabla \psi_0  + \nabla^2 \psi_0 \bigg ] -  \bigg (i \frac {q} {\hbar c} \vec{A} . \nabla \psi_0 \bigg )e^{i \frac {q} {\hbar c}\int^{\vec{r}} {\vec{A}(\vec{r'}).d\vec{r'}} } = e^{i \frac {q} {\hbar c}\int^{\vec{r}} {\vec{A}(\vec{r'}).d\vec{r'}} } \nabla^2 \psi_0 
\end{equation} 
Since we are considering a static magnetic field presently,  $\vec{A} $ is independent of time so that  we have from eqs.(18) and (17),
\begin{equation}
i \hbar {{\partial \psi}\over{\partial t}} = e^{i \frac {q} {\hbar c}\int^{\vec{r}} {\vec{A}(\vec{r'}).d\vec{r'}} } \bigg (i \hbar {{\partial \psi_0}\over{\partial t}} \bigg ) = e^{i \frac {q} {\hbar c}\int^{\vec{r}} {\vec{A}(\vec{r'}).d\vec{r'}} } \bigg ( - \frac {\hbar^2}{2m} \nabla ^2 \psi_0 \bigg )
\end{equation}
Because of the result we just obtained in eq.(23), the RHS of the above equation is simply,
\begin{equation}
- \frac {\hbar^2}{2m} \bigg (\nabla - i \frac {q} {\hbar c} \vec{A} \bigg ). \bigg (\nabla - i \frac {q} {\hbar c} \vec{A} \bigg ) \psi =   \frac {1} {2 m} \bigg (- i \hbar \nabla - \frac {q} {c}\  \vec{A}(\vec{r})\bigg ). \bigg (- i \hbar \nabla - \frac {q} {c}\  \vec{A}(\vec{r}) \bigg ) \psi
\end{equation}
This completes the proof that eq.(18) is indeed a solution of eq.(16) in field free regions.

\subsection{Aharonov-Bohm effect}

Now, comes the interesting part: Suppose we arrange a two-slit experiment in which a coherent beam of electrons ($q= -e = - 1.602 \times 10^{-19}$ coulomb) with a definite De Broglie wavelength $\lambda_e$ is incident at normal to a flat board with two very narrow rectangular slits between which  a thin and very long solenoid, whose axis is parallel to the length of the slits, has been  placed. 

When current passes through the solenoid, there is a non-zero magnetic field in  its interior. But outside the solenoid, $\vec {B} =0$. If $\lambda_e$ is much larger than the diameter of the solenoid, the electrons would not see the solenoid, leave alone its magnetic field inside. These conditions allow one to  use the wavefunction of eq.(18).

After passing through the slits at $\vec {r}_1 $  and $\vec {r}_2 $, the incident electron will hit the scintillation screen or a photographic plate at a point $\vec {r}_P $ like in the usual double-slit interference experiment. The wavefunction $\psi (\vec {r}_P, t)$ of the electron at $\vec {r}_P $ on the screen is given by the superposition of two wavefunctions corresponding to the propagation through the two slits at $\vec {r}_1 $  and $\vec {r}_2 $, respectively.  Using eq.(18), we have then,
\begin{equation}
\psi(\vec{r}_P,t)=e^{-\frac {i e}{\hbar c}\int^{\vec{r}_P}_{\vec{r}_0 \ C_1} {\vec{A}(\vec{r'}).d\vec{r'}}}\psi_1 (\vec{r}_P,t) + e^{-\frac {i e}{\hbar c}\int^{\vec{r}_P}_{\vec{r}_0 \ C_2} {\vec{A}(\vec{r'}).d\vec{r'}}}\psi_2 (\vec{r}_P,t)
\end{equation}
In the above equation $\vec {r}_0$ is an arbitrary point from where the electrons start and then pass through the slits at $ \vec {r}_1 $ and $ \vec {r}_2 $, respectively, and end up on the detector screen at  $ \vec {r}_P $. Because of the distinct slits, the paths $C_1$ and $C_2$ connecting $ \vec {r}_0$-$ \vec {r}_1 $-$ \vec {r}_P $ and $ \vec {r}_0$-$ \vec {r}_2 $-$ \vec {r}_P $, respectively, are chosen to be different so that they cannot be deformed into one another without cutting across the solenoid. These paths are purely mathematical, just as in the case of eq.(18), and do not represent the non-existent trajectories of electrons as discussed in section I.

The wavefunctions $\psi_1 $ and  $\psi_2 $, in eq.(26), represent the states of a free electron associated  with  slit 1 and slit 2, respectively, corresponding  to the case when no current flows through the solenoid (i.e. $\vec{B}=0$ everywhere). We can  always take one of the phase factors common in the RHS of eq.(26) so that,
\begin{equation}
\psi(\vec{r}_P,t)= e^{-\frac {i e}{\hbar c}\int^{\vec{r}_P}_{\vec{r}_0 \ C_2} {\vec{A}(\vec{r'}).d\vec{r'}}} [e^{-\frac {i e}{\hbar c}\int^{\vec{r}_P}_{\vec{r}_0 \ C_1} {\vec{A}(\vec{r'}).d\vec{r'}} + \frac {i e}{\hbar c}\int^{\vec{r}_P}_{\vec{r}_0 \ C_2}{\vec{A}(\vec{r'}).d\vec{r'}}} \psi_1 (\vec{r}_P,t) +  \psi_2 (\vec{r}_P,t)]
\end{equation}
Since a line-integral is always along a directed path, we can express the phase of $\psi_1$ as,
\begin{equation}
-\frac {i e}{\hbar c}\int^{\vec{r}_P}_{\vec{r}_0 \ C_1} {\vec{A}(\vec{r'}).d\vec{r'}} + \frac {i e}{\hbar c}\int^{\vec{r}_P}_{\vec{r}_0 \ C_2}{\vec{A}(\vec{r'}).d\vec{r'}} = -\frac {i e}{\hbar c}\int^{\vec{r}_P}_{\vec{r}_0 \ C_1} {\vec{A}(\vec{r'}).d\vec{r'}} -  \frac {i e}{\hbar c}\int^{\vec{r}_0}_{\vec{r}_P \ C_2}{\vec{A}(\vec{r'}).d\vec{r'}}  = -\frac {i e}{\hbar c}\oint_{C} {\vec{A}(\vec{r'}).d\vec{r'}}\ \ \ ,
\end{equation} 
where $C$ is the closed contour joining $ \vec {r}_0$-$ \vec {r}_1 $-$ \vec {r}_P $-$ \vec {r}_2$-$ \vec {r}_0$, that lies totally in a region where $\vec{B}=0$. But here comes the surprise.

From Stokes' theorem, we have the result,
\begin{equation}
\oint_{C} {\vec{A}(\vec{r'}).d\vec{r'}} = \int _S {\nabla \times \vec{A}(\vec{r'}). \vec {da}} = \int _S {\vec{B}(\vec{r'}). \vec {da}} \ \ \ ,
\end{equation}
where $S$ is the mathematical surface whose boundary is the contour $C$ and $\vec {da} $ is an infinitesimal area element on the surface $S$.

So, although the contour $C$ lies in the field-free region, the surface $S$ is pierced by the magnetic field due to the current carrying solenoid. Hence, the phase in eq.(28) is non-zero and is determined by the magnetic flux crossing the surface $S$. 

Applying the results of eqs.(28)and (29) in eq.(27), we obtain,
\begin{equation}
\psi(\vec{r}_P,t)= e^{-\frac {i e}{\hbar c}\int^{\vec{r}_P}_{\vec{r}_0 \ C_2} {\vec{A}(\vec{r'}).d\vec{r'}}} [e^{-\frac {i e}{\hbar c}\int_S {\vec{B}(\vec{r'}). \vec {da}}} \psi_1 (\vec{r}_P,t) +  \psi_2 (\vec{r}_P,t)]
\end{equation}
Therefore, the probability of finding an electron in an infinitesimal volume $dV$ at the position vector $ \vec {r}_P$ is given by,
\begin{equation}
P(\vec {r}_P, t) = \vert \psi(\vec{r}_P,t) \vert ^2 dV = \vert [e^{-\frac {i e}{\hbar c}\int_S {\vec{B}(\vec{r'}). \vec {da}}} \psi_1 (\vec{r}_P,t) +  \psi_2 (\vec{r}_P,t)]\vert ^2 dV
\end{equation}
This is the predicted Aharonov-Bohm (AB) effect according to which, if an experiment is performed first with no current in the solenoid, the observed interference pattern will be given by $\vert \psi_1 + \psi_2 \vert ^2 $. When the current is switched on, the fringes will shift according to eq.(31). This is completely a non-classical effect since the electrons never see the magnetic field (i.e. no Lorentz force due to $\vec{B}$) and yet the presence of $\vec{B}$ generates an additional  phase difference (eq.(28)) between the states associated with slits 1 and 2. 

AB effect has been observed  in several  experiments that range from thin magnetic whiskers  to miniscule, impenetrable toroidal magnets without leakage fields replacing the solenoid\cite{bec10,bec11,bec12,bec13}. The AB effect arises because of a combination of  quantum phase  and the peculiar topology of the region in which the charge particles move. The region accessible to the charges is not simply connected since the contour cannot be shrunk to a zero-size without leaving the field free region.  

The Aharonov-Bohm (AB) phase ${\frac { q}{\hbar c}\int^{\vec{r}}_{\vec{r}_0 \ C} {\vec{A}(\vec{r'}).d\vec{r'}}}$, appearing in eq.(18),  leads also to surprising  effects associated with superconductors (e.g. magnetic  flux quantisation) and with hypothetical magnetic monopoles (like, electric charge quantisation), as we shall see below. 

\subsection{Magnetic Flux Quantisation}

In a superconductor, not only  its electrical resistance vanishes below a critical temperature, it  displays perfect diamagnetism too so that $\vec {B}=0 $ inside its bulk. Such amazing phenomena arise due to the formation of Cooper pairs in superconductors. At sufficiently low temperatures,  a pair of  electrons with opposite spin states can become a weakly held  bound system, referred to as the Cooper pair, by exchanging virtual phonons \cite{bec14, bec15}. Phonons are basically the propagating quanta of elastic waves (or, lattice waves) in a substance.

An intuitive way to understand the mechanism behind the formation of Cooper pairs goes as follows.  Although electrons repel each other electrostatically, an electron moving in a solid endowed with a lattice structure attracts the positively charged ions present in the lattice sites. As the ions crowd towards an  electron whizzing by, they deform the lattice structure giving rise to a tiny excess of positive charge density in a small region for a short duration. Another approaching electron gets attracted electrostatically towards this region because of the net positive charge density created by the previous electron. As a result, there is an effective attractive force mediated by lattice deformation between the two electrons separated in time and space.

When the temperature of the system is sufficiently low,  thermal energy associated with random motion of electrons and ions in the substance is unable, on an average, to break apart these loosely held pair of electrons. Because of the opposite spin states, a Cooper pair has 0 spin and hence is a boson. At low temperatures, the thermal De Broglie wavelength of such Cooper pairs can be larger than the average separation between mutual pairs, leading to Bose-Einstein condensation (BEC)  so that all the bosonic pairs occupy essentially the same quantum level \cite{bec14}. 

The Cooper pairs, each having an electric charge $-2 e$, responsible for the counter-intuitive behaviour of superconductors can therefore be described by a macroscopic wavefunction $\Psi (\vec{r}, t)$. In normal conductors, electrical resistance arise due to random scattering of electrons by phonons.  One may  then provide the following physical description to explain the vanishing of  resistance in superconductors.

 Since a sizable fraction of electrons form Cooper pairs and undergo BEC, a majority of the pairs occupy a single, low energy quantum level corresponding to a macroscopic wavefunction $\Psi_0 (\vec{r}, t)$ that  constitute a flowing current. At low temperatures, the randomly moving phonons do not have sufficient energy to excite these pairs, and hence the average current, associated with the state $\Psi_0 (\vec{r}, t)$,  flows in an uninterrupted manner. After all, in an atom too electrons can have quantum states with non-zero orbital angular momentum, leading to effectively a tiny electrical current that does not get dissipated unless high energy photons excite these electrons. 
 
 Superconductors are perfect diamagnets. When a solid ball of a superconductor is placed in an external magnetic field, the $\vec {B} $ field in the interior is zero. This feature is called the {\bf Meissner effect} and  can be understood intuitively as follows. According to   Lenz law, when a conductor approaches a region having $\vec {B}\neq 0 $, currents flow  in such way as to oppose the rise of the field in the conductor. Similarly, surface currents get generated in a  superconductor that is subjected to an external $\vec {B} $ so as to nullify the net $\vec {B} $ field inside the superconductor. Since there is no electrical resistance, the surface current flows indefinitely,  maintaining a net $\vec {B}=0 $ in the interior. 
 
Consider now a superconductor in the shape of a dough-nut  being introduced in a static,  external magnetic field.  Of course, as argued in the previous paragraph, we expect $\vec {B} $ to be zero in the superconducting material. But there is a hole  too, because of the dough-nut shape. What would  the field be in this empty region `ringed' by a thick loop of superconductor?

Clearly, this is a case in which we can immediately apply the AB-phase of eq.(18), as the Cooper pairs do not directly see any $\vec {B}$. \cite{bec16} If in the absence of  magnetic field the macroscopic wavefunction describing the pairs is $\Psi_0 $ then, as discussed in section IV, when the field is switched on, the wavefunction $\Psi (\vec{r})$, at any position vector $\vec {r}$ within the superconductor, is  given by,
\begin{equation}
\Psi (\vec{r})=   e^{-\frac {2i e}{\hbar c}\int^{\vec{r}}_{\vec{r}_0 \ C} {\vec{A}(\vec{r'}).d\vec{r'}}}  \Psi_0 
\end{equation}
where $\vec {r}_0$ is some point in the superconducting substance, and $C$ is a directed path lying within the material and connecting $\vec {r}_0$ to $\vec {r}$. Again, the phase in the above equation is path independent since $\nabla \times \vec {A} (\vec{r}) =0$ inside the dough-nut.  The factor $2$  in the phase of eq.(32) appears because of the fact that the electric charge of a Cooper pair is $ - 2 e$. 

Suppose we start from a position $\vec {r}_1$, where the wavefunction is $\Psi (\vec{r}_1)$  as dictated by eq.(32), and move to any point $\vec{r}_P$ being all the while inside  the  substance. Then, the wavefunction at $\vec{r}_P$  will be simply $\Psi (\vec{r}_P) $, again due to eq.(32),
\begin{equation}
\Psi (\vec{r}_P)=   e^{-\frac {2i e}{\hbar c}\int^{\vec{r}_P}_{\vec{r}_0 \ C} {\vec{A}(\vec{r'}).d\vec{r'}}}  \Psi_0= e^{-\frac {2i e}{\hbar c}\int^{\vec{r}_P}_{\vec{r}_1 \ C_2}{\vec{A}(\vec{r'}).d\vec{r'}}}  e^{-\frac {2i e}{\hbar c}\int^{\vec{r}_1}_{\vec{r}_0 \ C_1} {\vec{A}(\vec{r'}).d\vec{r'}}} \Psi_0= e^{-\frac {2i e}{\hbar c}\int^{\vec{r}_P}_{\vec{r}_1 \ C_2}{\vec{A}(\vec{r'}).d\vec{r'}}}\Psi (\vec{r}_1) 
\end{equation}
In the above equation, we have just broken the path $C$ from $\vec{r}_0$ to $\vec{r}_P$ into two segments,  $C_1$, from $\vec{r}_0$ to $\vec{r}_1$ and then $C_2$, $\vec{r}_1$ to $\vec{r}_P$.

 Now comes the magic: As we move continuously from $\vec {r}_1$ and then return to the point  $\vec {r}_1$  by tracing a closed contour $C_{cl}$ around the hole of the dough-nut, the wavefunction according to eq.(33) becomes,
\begin{equation}
\Psi^\prime (\vec{r_1})=   e^{-\frac {2i e}{\hbar c}\oint {\vec{A}(\vec{r'}).d\vec{r'}}}  \Psi_ (\vec{r}_1)= e^{-\frac {2i e}{\hbar c}\int_S {\vec{B}(\vec{r}).d\vec{a}}}  \Psi_ (\vec{r}_1)
\end{equation} 
In the above equation, $S$ is the surface bounded by the closed contour $C_{cl}$.  We have used the Stokes' theorem (eq.(29)) in  eq.(34) to arrive at the final result.

But the wavefunction must be single-valued, so that $\Psi^\prime (\vec{r_1})= \Psi (\vec{r_1})$,  implying that the phase,
\begin{equation}
\frac {2 e}{\hbar c}\oint {\vec{A}(\vec{r'}).d\vec{r'}}=\frac {2 e}{\hbar c}\int_S {\vec{B}(\vec{r}).d\vec{a}}= 2 \pi n\ \ \ n=0,1,2,....
\end{equation} 
We have, therefore, obtained a remarkable result that any superconductor that has  a hole in it traps  magnetic field  in the empty region in a way that the magnetic flux is quantised in steps of $\frac {2\pi \hbar c n} {2e}$.  It is to be noted that the presence of  $2 e$ in the denominator confirms the crucial role that Cooper pairs play in superconductivity. Flux quantisation  has been  verified in a large body of experiments.

 Again, it is the interplay between the AB phase and the superconductor not being simply connected  that entails  magnetic flux quantisation.  Such quantised fluxes, $\frac {\pi \hbar c n} {e}$, $n=0,1,2,...$,  have also been observed in the case of high $T_C$ superconductors, substantiating  the idea that  pairs of electrons (i.e. Cooper pairs) with charge $-2 e$ are involved in this class of superconductors too.
 
\subsection{Magnetic monopoles and Charge Quantisation} 

Now, we take up the issue of magnetic monopoles. It is a common observation that if one breaks a bar magnet into two, one does not get a north  magnetic pole and a south magnetic pole. Instead one ends up having two smaller pieces of bar magnets. This is not surprising because the magnetism exhibited by ferromagnetic materials is essentially due to their atoms carrying tiny magnetic dipole moments brought upon   by their constituent charge particles having intrinsic spin angular momentum.    

On the other hand, if magnetic monopoles were to exist, Maxwell equations would be  more symmetric, with electric and magnetic fields having the same status \cite{bec17, bec18, bec19}. From a pioneering study by Dirac in 1931 it ensues that the presence of magnetic monopoles  can also explain why the observed  electric charge  is always quantised  \cite{bec17}.

 However,  if monopoles exist then 
 the 2-nd Maxwell equation gets modified to $\nabla . \vec{B} = 4 \pi \rho_{mono} (\vec{r})$, where $\rho_{mono} (\vec{r})$ is the density of magnetic charge. For instance, if one has a point magnetic monopole with magnetic charge $g$, sitting at the origin $\vec{r}=0$, then $\rho_{mono}= g \  \delta^3 (\vec{r})$ so that,
 \begin{equation}
 \nabla . \vec{B} = 4 \pi \rho_{mono} (\vec{r}) = g \   \delta^3 (\vec{r}) \Rightarrow  \vec{B}= \frac{g} {r^3}\  \vec{r} 
\end{equation}  
 
Of course, if $\rho_{mono} \neq 0$, one cannot have an `everywhere well defined and smooth' vector potential $\vec {A}$ since for any such non-singular vector field, $\nabla . \vec{B} = \nabla . (\nabla \times \vec{A}) = 0$.  However, Dirac had noted that one can construct singular vector potentials such that $\nabla . \vec{B} \neq 0$.

For a point monopole with magnetic charge $g$, adopting a spherical polar coordinate system,  one may construct the vector potential $\vec{A}(\vec{r})$ to be either \cite{bec20},
\begin{equation}
A_r= 0 = A_\theta  \ \ \ \mbox{and}\ \ \ A_\phi = \frac { g \ (1-\cos{\theta})} {r \ \sin{\theta}} \ \ ,
\end{equation}
or,
\begin{equation}
A_r= 0 = A_\theta  \ \ \ \mbox{and}\ \ \ A_\phi = - \frac { g \ (1+\cos{\theta})} {r \ \sin{\theta}} \ \ ,
\end{equation}
From eqs.(37) it is obvious that $\vec{A}(\vec{r})$ is singular at every point on the negative z-axis since $A_\phi $ is infinity when $\theta = \pi$. On the other hand, the vector potential given by eq.(38)    is singular at every point on the positive z-axis as $A_\phi $ blows up when $\theta = 0$. However, both eqs.(37) as well as (38) result in the same magnetic field (eq.(36))  when the curl is taken. 

The lines of singularity (often referred to as Dirac strings) are therefore  mathematical artifacts, for the magnetic field derived from the singular $\vec{A}$ of either eq.(37) or of eq.(38) are not only identical but are also well behaved everywhere except at the origin (which is expected since we have considered a point magnetic charge).

Physically, one can think of the magnetic monopole as one end of an infinitely long and thin, current carrying solenoid. This solenoid, which in reality does not exist physically,  then is the Dirac string.    

Suppose we consider the double-slit experiment with a coherent beam of charge particles with charge $q$, similar to that discussed in the subsection A, but this time without a solenoid but instead with a magnetic monopole of charge $g$ located at the origin that is  very, very far away from the slits. That is,  $\vert \vec {r}_1 \vert $, $\vert \vec {r}_2 \vert $, $\vert \vec {r}_P \vert $, etc. are so  large that the magnetic Lorentz force due to the monopole on a charge particle is negligible compared to all other forces acting on it in the experimental set up. This condition ensures that we can apply eq.(18) in the present circumstance too. 

 Then, from eq.(31),  the probability  of detecting  an incident particle (after it has passed through the slits) in an infinitesimal volume element at $\vec{r}_P$ is given by,
\begin{equation}
P(\vec {r}_P, t) = \vert \psi(\vec{r}_P,t) \vert ^2 dV = \vert [e^{\frac {i q}{\hbar c}\int_S {\vec{B}(\vec{r'}). \vec {da'}}} \psi_1 (\vec{r}_P,t) +  \psi_2 (\vec{r}_P,t)]\vert ^2 dV
\end{equation}
where $S$ is the surface bounded by the closed contour  joining $ \vec {r}_0$-$ \vec {r}_1 $-$ \vec {r}_P $-$ \vec {r}_2$-$ \vec {r}_0$ like in subsection A. 

Now, if the  Dirac string does not pierce through $S$ then $\frac {q}{\hbar c}\int_S {\vec{B}(\vec{r'}). \vec {da}} \cong 0$,  as the magnetic field is negligibly small where the experiment is being conducted. Hence, the interference fringe pattern does not get affected by the monopole. 

 But if the Dirac string  passes through the surface $S$, the phase difference is given by,
\begin{equation}
{\frac {q}{\hbar c}\int_S {\vec{B}(\vec{r'}). \vec {da'}}} = \frac {q}{\hbar c} \oint_{C}  {\vec{A}(\vec{r'}).d\vec{r'}} =  
\end{equation}
\begin{equation}
= \frac {q}{\hbar c} \oint_{C'}  {\vec{A}(\vec{r'}).d\vec{r'}}= \frac {q}{\hbar c} \int_{S'} {\vec {B}.\vec {da'}}= \frac {q}{\hbar c}\bigg ( \frac {g}{r^2}\bigg )  4 \pi r^2 =
\frac {4 \pi q g}{\hbar c} \ \ \ ,
\end{equation}
where $S'$ is a spherical surface around the magnetic monopole (i.e. the  end of the solenoid) not containing the point of intersection with the Dirac string so that it is bounded by an infinitesimally  tiny contour $C'$ going around the Dirac string. Eq.(41) follows from eq.(40) because of the reason that the closed contour $C$ forming the boundary of the surface $S$ can be shrunk to $C'$ without changing the value of the integral. 

One can understand eqs.(40) and (41)  physically too by taking  the Dirac string to be a  semi-infinite, thin solenoid, as mentioned before.  The magnetic flux, $\int_{S} {\vec {B}.\vec {da'}}$, is then 
 equal to the flux across the cross-section of the solenoid, which is simply  the magnetic flux, $\int_{S'} {\vec{B}(\vec{r'}). \vec {da'}}$, due to the field of the monopole that  crosses the spherical surface $S'$ since $\nabla . \vec {B}=0$.
 
  But the Dirac string (equivalently, the semi-infinite solenoid) is an artifact, and its presence or absence should not affect the interference pattern. In other words, whether the phase difference in eq.(39) is zero or $\frac {4 \pi q g}{\hbar c}$ is immaterial to the interference experiment. This entails that the phase difference,
\begin{equation}
\frac {4 \pi q\  g}{\hbar c}= 2 \pi n \ \ \ \Rightarrow  q \ g = \frac {1} {2} n \hbar c, \ n=0,1,2,.... 
\end{equation}

Eq.(42) implies that the presence of a single magnetic monopole anywhere in the universe would lead to electric charge be quantised, $q =  \frac {1} {2 g} n \hbar c$, $ n=0,1,2,... $.

\section* {Summary}
Gauge symmetry dictates how the electromagnetic  potentials ought to couple with the wavefunctions describing charge particles. Making use of this, one can determine the wavefunction $\psi$ of a charge particle, when magnetic field is switched on, in the regions where $\vec {B}=0$  if one already has a  solution $\psi_0$ of the Schrodinger equation corresponding to a free particle. The wavefunction $\psi $ is obtained by  simply multiplying the Aharonov-Bohm phase factor to $\psi_0$ so that $\psi = [\exp{(\frac {i q}{\hbar c}\int^{\vec{r}}_{\vec{r}_0 \ C} {\vec{A}(\vec{r'}).d\vec{r'}})} ] \psi_0$ describes the state of the particle in regions free of electromagnetic fields.

In the preceding sections,  implications of the Aharonov-Bohm phase factor, $[\exp{(\frac {i q}{\hbar c}\int^{\vec{r}}_{\vec{r}_0 \ C} {\vec{A}(\vec{r'}).d\vec{r'}})} ]$ were discussed for  three distinct physical situations -  when applied to the double-slit interference experiment,  in the superconducting ring problem and in the case of a test charge interacting with a  magnetic monopole. In all these cases, charge particles were confined to regions where $\vec {B}=0$  so that the   closed contours associated with the Aharonov-Bohm phase of the wavefunction could not be shrunk to a point. In other words, the AB phase along with the `holes' of the non-simply connected regions corresponding to the  $\psi=0$ and $\vec {B}\neq 0$ sites, led to the novel features like AB effect, flux quantization and Dirac charge quantization, respectively.

\begin{acknowledgments} 
I thank Prof. Sudhendu Rai Choudhury and Prof. Kamal Datta for their useful comments in the past  while discussing foundations of quantum theory.
\end{acknowledgments}


\end{document}